\documentclass{article}
\usepackage{float}
\usepackage{epsfig}
\usepackage{xcolor}
\usepackage{soul}
\usepackage{amsmath}
\usepackage{graphicx}
\usepackage[colorlinks=true,
            linkcolor=blue,
            citecolor=red,
            urlcolor=cyan,
            pdfborder={0 0 0}, 
            backref=true] 
{hyperref}

\newcommand{\redcite}[1]{\textcolor{red}{\cite{#1}}}

\makeatletter
\def\tagform@#1{\maketag@@@{\color{red}(#1)}}
\makeatother
\begin{document}
\title{\bf Rotating black holes in the Hernquist galactic halo and its accretion disk luminosity}
\author{{ Malihe Heydari-Fard$^{1}$\thanks{Electronic address: heydarifard@qom.ac.ir} and Mohaddese Heydari-Fard$^{1}$ \thanks{Electronic address: mo.heydarifard@qom.ac.ir} }\\ {\small \emph{$^{1}$ Department of Physics, The University of Qom, 3716146611, Qom, Iran}}}

\maketitle

\begin{abstract}
Static, spherically symmetric black holes immersed in a dark matter halo with a Hernquist-type density profile have been derived by Cardoso et. al. in Ref. \redcite{Cardoso:2021wlq}. Using the Newman-Janis algorithm, we construct the metric for a stationary and axially symmetric rotating black hole in this environment. Then, we obtain the electromagnetic properties of thin accretion disks around such rotating black holes by utilizing the steady-state Novikov-Thorne model, and study the effects of spin parameter and halo compactness parameter on the disk properties. Finally, by comparison the results of the rotating Cardoso black hole with that of Kerr black hole in the absence of dark matter, we find that the presence of dark matter can not significantly affect the disk properties and thus for astrophysical black holes with large spin parameter, the distinction of rotating Cardoso black holes becomes more difficult than the Kerr black hole.
\vspace{5mm}
\vspace{1mm}
\end{abstract}

\section{Introduction}
In recent years, the study of strong gravity regimes has undergone profound revolution. With the discovery of gravitational waves from binary black hole (BH) mergers \textcolor{red}{\cite{1}} and the observation of images of supermassive BHs, M87* and SgrA*, at the centers of the giant elliptical galaxy M87 and the Milky Way \textcolor{red}{\cite{2, 3}}, this area of physics—from the field mostly disconnected from observations—has become one of the most vibrant and active research fields, providing access to the most mysterious massive objects such as BHs.

On the other hand, observational data including the rotation curves of spiral galaxies, the cosmic microwave background radiation, and gravitational lensing confirms the existence of dark matter in the cosmos. Based on this data, a series of dark matter models for its study have been proposed by astronomers. From the phenomenological aspect, the radial distribution of dark matter halos is often described using cuspy or cored profiles. N-body simulations of cold dark matter models typically predict cuspy profiles such as the Navarro–Frenk–White (NFW) profile \textcolor{red}{\cite{Navarro:1995iw}}. Matos et. al. obtained the space-time geometry of dark matter halo for NFW density profile \redcite{Matos:2003nb}. However, observations of low-surface-brightness and dwarf galaxies often predict cored profiles with an approximately constant central density, such as the Burkert density profile \textcolor{red}{\cite{Burkert:1995yz, Salucci:2000ps}}. Using the dark matter halos of seven dwarf spiral galaxies, Burkert obtained the density of universal rotation curve profile \textcolor{red}{\cite{Burkert:1995yz}}.

It is believed that supermassive BHs reside in the centers of many galaxies. Moreover, the cosmos is filled with matter, which provides an opportunity to test the evolution of these BHs in such environments. In fact, according to the standard model, the matter content of our universe is composed of $27\%$  of invisible dark matter, which may accumulate at galactic centers and close to BHs, affecting the dynamics of compact binary systems and the propagation of gravitational or electromagnetic waves. Therefore, it  is important to study the supermassive BHs immersed in dark matter halos.

Using Newtonian approximations, Gondolo et al. derived the dark matter density profile around the BH at the center of the Milky Way galaxy \textcolor{red}{\cite{Gondolo:1999ef}}. To generalize Newtonian approximations to relativistic calculations, Sadeghian et al. used the exact Schwarzschild geometry and showed that the density peaks near $r \sim 2 r_s$ have a steep cutoff at $r = 2 r_s$, below which the dark matter density profile vanishes \textcolor{red}{\cite{Sadeghian:2013laa}}. Both Newtonian and relativistic analyses indicate that, in the presence of a galactic center BH, the density profile close to its horizon approaches zero. The dark matter profile creates a spike close to the horizon, whose length-scale is determined by the BH mass $M_{BH}$. Considering different dark matter halo profiles, the space-time geometry of a BH immersed in a dark matter halo has been extensively studied. For example, Xu et al. derived the space-time geometry of a BH immersed in a dark matter halo using the NFW profile \redcite{Xu:2018wow}--\redcite{Xu:2020jpv}. The analysis of the dark matter spike effects on the BH at the Milky Way’s center has been done in \redcite{Nampalliwar:2021tyz}. Similarly, the influence of the Einasto dark matter profile on the central BH has been explored in Ref. \redcite{Liu:2023oab}. For BHs surrounded by generic dark matter profiles (Hernquist, NFW, and Einasto), Cardoso et al. by introducing a numerical method \redcite{Figueiredo:2023gas} investigated the effects of these environments on null and time-like geodesics, as well as on gravitational wave propagation. Furthermore, Cardoso et al., through an interesting and novel approach using the "Einstein cluster" model, provided a relativistic model for BHs immersed in dark matter halos with a Hernquist-type density and anisotropic pressure, examining the galactic environment’s effects on gravitational waves \redcite{Cardoso:2021wlq}. Following Cardoso's method, the authors in \redcite{Konoplya:2022hbl} have presented a numerical framework for computing the metric of static spherically symmetric BHs in environments with NFW, Burkert, and Taylor-Silk density profiles. Employing a similar approach, Shen et al. considered five analytical dark matter halo models near supermassive BHs and provided the corresponding BH solutions \redcite{Shen:2023erj}. The influence of the Einasto profile on the supermassive BH in M87 has also been investigated in Ref. \redcite{Chowdhury:2025tpt}.

The fact that real BHs reside in galactic environments (dark matter, stars, and gas), on one hand, and the consistency between the shadow of a rotating Kerr BH and Event Horizon Telescope images, on the other hand, motivates investigating the optical appearance of rotating BHs immersed in dark matter halos. Therefore, in the present work, we study the accretion process in thin disks around rotating BHs in the Hernquist galactic halo and investigate the effects of dark matter on their properties. Note that our study differs from Ref. \redcite{Balali:2024mtt}, where the authors constructed the slowly rotating BH solution in the Hernquist-type density and it appears that the results obtained in the slow-rotation limit are not distinguishable from the static case. Moreover, authors in \redcite{Balali:2024mtt} studied photon trajectories on the null geodesics and the BH shadow, while we will consider the geodesic motion of test particles in thin disks around a central BH.

The paper is organized as follows: In Section~\ref{2-disk}, we give a brief review of the thin accretion disk model. We present the Schwarzschild-like BH solution immersed in a dark matter halo with Hernquist-type density profile in subsection~\ref{3-1-static}. Then, in subsection~\ref{3-2-rotating}, we construct the rotating counterpart of this static solution, namely, the rotating BHs at the center of Hernquist-type density distribution using the Newman-Janis algorithm, and also discuss the electromagnetic properties of accretion disks around such BHs in subsection~\ref{3-3-disk}. The paper ends with concluding remarks in Section~\ref{4}.

\section{A review of the thin accretion disk model}
\label{2-disk}
In this section we first present a brief review of the equatorial circular orbits in a general stationary and axisymmetric space-time and then represent
the main equations that describe the electromagnetic radiation properties of thin relativistic accretion disks.

The general form of the metric for a stationary and axisymmetric space-times is given by
\begin{equation}
ds^2=g_{tt}dt^2+2g_{t\phi}dtd\phi+g_{rr}dr^2+g_{\theta\theta}d\theta^2+g_{\phi\phi}d\phi^2,
\label{1}
\end{equation}
where all metric components depend only on $r$ and $\theta$ coordinates. In such space-times, the symmetries associated with the Killing vectors $\partial_{t}$ and $\partial_{\phi}$ yield two constants of motion for test particles, the energy per unit rest-mass $\tilde{E}$, and the angular momentum per unit rest-mass, $\tilde{L}$, as
\begin{equation}
g_{tt}\dot{t}+g_{t\phi}\dot{\phi} = -\tilde{E},
\label{1}
\end{equation}
\begin{equation}
g_{t\phi}\dot{t}+g_{\phi\phi}\dot{\phi} = \tilde{L},
\label{2}
\end{equation}
where a dot denotes derivative with respect to the parameter $\tau$. Now, we can find $t$ and $\phi$ components of the $4$-velocity $\dot{x}^{\mu}$ as
\begin{equation}
\dot{t}=\frac{\tilde{E}g_{\phi\phi}+\tilde{L}g_{t\phi}}{g_{t\phi}^2-g_{tt}g_{\phi\phi}},
\label{4}
\end{equation}
\begin{equation}
\dot{\phi}=-\frac{\tilde{E}g_{t\phi}+\tilde{L}g_{tt}}{g_{t\phi}^2-g_{tt}g_{\phi\phi}}.
\label{5}
\end{equation}
By using the normalization condition for the four-velocity, $g_{\mu\nu}\dot{x}^{\mu}\dot{x}^{\nu} = -1$, we obtain
\begin{equation}
g_{rr} \dot{r}^2+g_{\theta\theta}\dot{\theta}^2=V_{\rm eff}(r,\theta),
\label{6}
\end{equation}
with the effective potential defined as
\begin{equation}
V_{\rm eff}(r,\theta)=-1+\frac{\tilde{E}^2g_{\phi\phi}+2\tilde{E}\tilde{L}g_{t\phi}+\tilde{L}^2g_{tt}}{g_{t\phi}^2-g_{tt}g_{\phi\phi}}.
\label{7}
\end{equation}
Applying the conditions for equatorial circular orbits, $\dot{r}=\dot{\theta}=\ddot{r} = 0$, to the radial geodesic equation yields the following expression for the orbital angular velocity $\Omega=\dot{\phi}/\dot{t}$  \cite{Bambi}
\begin{equation}
\Omega_{\pm}=\frac{-g_{t\phi,r}\pm\sqrt{(g_{t\phi,r})^2-g_{tt,r}g_{\phi\phi,r}}}{g_{\phi\phi,r}},
\label{8}
\end{equation}
here, plus and minus signs correspond to co-rotating and counter-rotating orbits, respectively. Then, use of the normalization condition, $g_{\mu\nu}\dot{x}^{\mu}\dot{x}^{\nu} = -1$, and equations \textcolor{red}{(\ref{1})} and \textcolor{red}{(\ref{2})} leads to the explicit expressions for the $\tilde{E}$ and $\tilde{L}$ of a particle on a circular orbit around a central massive object
\begin{equation}
{\tilde{E}}=-\frac{g_{tt}+g_{t\phi}\Omega}{\sqrt{-g_{tt}-2g_{t\phi}\Omega-g_{\phi\phi}\Omega^2}},
\label{9}
\end{equation}
\begin{equation}
{\tilde{L}}=\frac{g_{t\phi}+g_{\phi\phi}\Omega}{\sqrt{-g_{tt}-2g_{t\phi}\Omega-g_{\phi\phi}\Omega^2}}.
\label{10}
\end{equation}
For test particles orbiting a central massive object, the innermost stable circular orbit (ISCO) is defined as
\begin{equation}
V_{\rm eff,rr}\mid_{r=r_{\rm isco}}=\frac{1}{g_{t\phi}^2-g_{tt}g_{t\phi}}\left[{\tilde{E}^2g_{\phi\phi,rr}}
+2\tilde{E}\tilde{L}g_{t\phi,rr}+{\tilde{L}^2g_{tt,rr}}-\left(g_{t\phi}^2-g_{tt}g_{\phi\phi}\right)_{,rr}\right]\mid_{r=r_{\rm isco}}=0.
\label{11}
\end{equation}

Our analysis of accretion disks relies on the Novikov-Thorne model which describes the geometrically thin and optically thick accretion
disks around BHs. We require key observable parameters like the energy flux $F(r)$, temperature $T(r)$, luminosity spectra $L(\nu)$, and efficiency $\epsilon$. The model's assumptions are as follows: the space-time is stationary, axisymmetric, and asymptotically flat; the disk is geometrically thin and lies in the equatorial plane, $\theta=\pi/2$; its self-gravity is negligible; and its inner edge is located at $r_{\rm isco}$ since equatorial circular orbits become unstable for $r < r_{\rm isco}$. Furthermore, the disk is assumed to be in local thermal equilibrium, so its emission follows a black body law
\begin{equation}
F(r)=\sigma_{\rm SB} T(r)^4,\label{12}
\end{equation}
where $\sigma_{\rm SB}=5.67\times10^{-5}\rm erg$ $\rm s^{-1} cm^{-2} K^{-4}$ is the Stefan-Boltzmann constant and the radiant energy flux over the disk surface can be expressed as
\cite{Novikov}--\cite{Page}
\begin{equation}
F(r)=-\frac{\dot{M}_{0}\Omega_{,r}}{4\pi \sqrt{-g}\left(\tilde{E}-\Omega \tilde{L}\right)^2}\int^r_{r_{\rm isco}}\left(\tilde{E}-\Omega \tilde{L}\right) \tilde{L}_{,r}dr,
\label{13}
\end{equation}
where $\dot{M}_{0}$ is the mass accretion rate, which for the Novikov-Thorne model is constant.

Also, the observed luminosity of a thin disk has a red-shifted black body spectrum given by
\cite{Torres}
\begin{equation}
L(\nu)=4\pi d^2 I(\nu)=\frac{8\pi h \cos\gamma}{c^2}\int_{r_{\rm in}}^{r_{\rm out}}\int_0^{2\pi}\frac{ \nu_e^3 r dr d\phi}{\exp{[\frac{h\nu_e}{k_{\rm B} T}]}-1
},\label{14}
\end{equation}
where $d$ is the distance to the source, $\gamma$ is the disk inclination angle and $h$ and $k_{\rm B}$ are the Planck and Boltzmann constants, respectively. The emitted frequency at disk, $\nu_{e} = \nu (1+z)$, is related to the observed frequency $\nu$, through the redshift factor $z$ determined by the space-time metric and orbital motion as
\begin{equation}
1+z=\frac{1+\Omega r\sin\phi\sin\gamma}{\sqrt{-g_{tt}-2g_{t\phi}\Omega-g_{\phi\phi}\Omega^2}}.\label{15}
\end{equation}
Another important quantity is the radiative efficiency, which measures the BH's capability to convert accreted rest mass into electromagnetic radiation. When photon capture by the BH is negligible, the Novikov-Thorne efficiency is given by \cite{Thorne}
\begin{equation}
\epsilon=1-\tilde{E}_{\rm isco},\label{16}
\end{equation}
where $\tilde{E}_{\rm isco}$ is the specific energy of test particles at the ISCO radius.

\section{Thin accretion disk properties around rotating Cardoso BHs}
\label{3}
\subsection{Schwarzschild-like BHs with Hernquist density}
\label{3-1-static}
In Ref. \redcite{Cardoso:2021wlq}, Cardoso et. al. have derived galactic BHs immersed in a dark halo by considering an anisotropic dark matter fluid with tangential pressure $p_t$ and zero radial pressure $p_r=0$. The line-element for this Schwarzschild-like solution (static Cardoso BH) is given by
\begin{align}
ds^{2} = -\left(1-\frac{2M_{\mathrm{BH}}}{r}\right)e^{\gamma(r)}dt^{2}
       + \frac{dr^{2}}{\left[1-\frac{2M_{\mathrm{BH}}}{r}-\frac{2Mr}{(a_{0}+r)^{2}}\left(1-\frac{2M_{\mathrm{BH}}}{r}\right)^{2}\right]}
       + r^{2}\left(d\theta^{2}+\sin^{2}\theta d\phi^{2}\right),\label{17}
\end{align}
where
\begin{equation}
\gamma(r)=-\pi\sqrt{\frac{M}{\xi}}+2\sqrt{\frac{M}{\xi}}\;\mathrm{arctan}\left(\frac{r+a_{0}-M}{\sqrt{M\xi}}\right),\label{18}
\end{equation}
and
$$
\xi=2a_{0}-M+4M_{\mathrm{BH}},\label{19}
$$
here $M$ is the mass of the halo and $a_0$ is the typical length-scale of the galaxy. We have studied the electromagnetic radiation properties of thin accretion discs around static Cardoso BH (17) in Ref. \redcite{Heydari-Fard:2024wgu}. In the following, we will aim to extend this static solution to the case of rotation and investigate the accretion disk properties around rotating BHs immersed in a dark halo with a Hernquist-type density profile.

\subsection{Kerr-like BHs with Hernquist density}
\label{3-2-rotating}
\begin{figure}[H]
\centering
\includegraphics[width=3.0in]{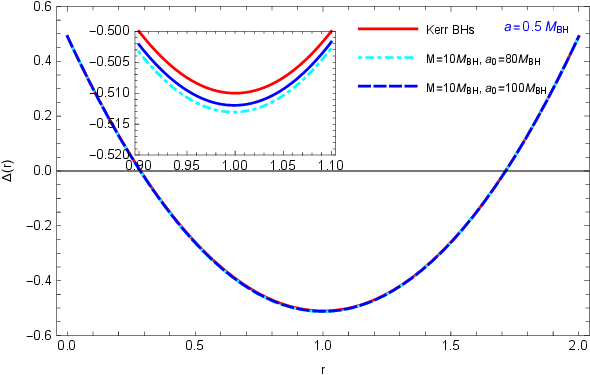}
\includegraphics[width=3.0in]{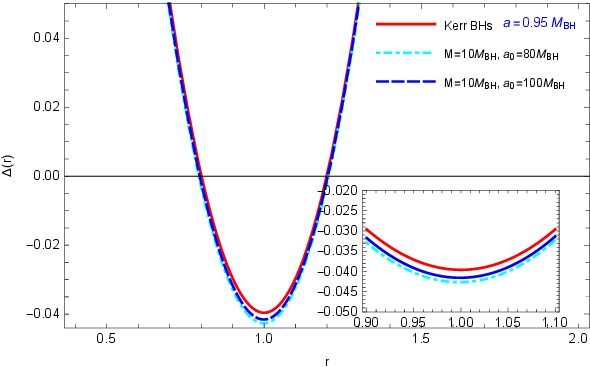}
\caption{\footnotesize The behavior of horizons as a function of $r$ for different values of ${\cal C}$ and rotation parameter $a=0.5M_{\rm BH}$ (left panel), and $a=0.95M_{\rm BH}$ (right panel), respectively. The solid curve corresponds to the Kerr BH without dark matter.}
\label{horizon}
\end{figure}

In this section, we focus on deriving a rotating BH solution relevant to galactic nuclei environments. Authors in Ref. \redcite{Liu:2023oab} by starting a general static spherically symmetric metric as
\begin{align}
ds^{2} = -F(r) dt^{2} + \frac{1}{G(r)} dr^{2} + H^2(r)\left(d\theta^{2}+\sin^{2}\theta d\phi^{2}\right),\label{20}
\end{align}
extended the Schwarzschild-like metric to the Kerr-like metric by using the Newman-Janis algorithm. The final metric in the Boyer-Lindquist coordinates is given by
\begin{equation}
\begin{split}
ds^2 &= -\frac{(GH + a^2 \cos^2 \theta) \Sigma}{(K + a^2 \cos^2 \theta)^2} dt^2
       + \frac{\Sigma}{GH + a^2} dr^2 - 2a \sin^2 \theta \left[ \frac{K - GH}{(K + a^2 \cos^2 \theta)^2} \right] \Sigma \, dt \, d\phi \\
     &\quad + \Sigma \, d\theta^2 + \Sigma \sin^2 \theta \left[ 1 + \frac{a^2 \sin^2 \theta (2K - GH + a^2 \cos^2 \theta)}{(K + a^2 \cos^2 \theta)^2} \right] d\phi^2.\label{21}
\end{split}
\end{equation}
The above metric describes the rotating BHs immersed in a dark halo with a Hernquist-type density profile (rotating Cardoso BH) if we define $F$, $G$ and $H$ as
\begin{align}
F(r) = \left(1-\frac{2M_{\mathrm{BH}}}{r}\right)e^{\gamma(r)},\label{22}
\end{align}
\begin{align}
G(r) = {\left[1-\frac{2M_{\mathrm{BH}}}{r}-\frac{2Mr}{(a_{0}+r)^{2}}\left(1-\frac{2M_{\mathrm{BH}}}{r}\right)^{2}\right]},\label{23}
\end{align}
$$H(r) = r^2,$$
with
\begin{align}
K(r) = H(r)\sqrt{\frac{G(r)}{F(r)}},\label{24}
\end{align}
and
\begin{align}
\Sigma = r^2+a^2 cos^2\theta,\label{25}
\end{align}
\begin{align}
\Delta = r^2 G(r)+a^2,\label{26}
\end{align}
which $\gamma(r)$ was introduced in equation \textcolor{red}{(\ref{18})}. Here, it should be noted that the authors in Ref. \redcite{Balali:2024mtt} extended the static Cardos BH to the case of slowly rotating BH which does not include an astrophysical BH with a large spin parameter, whereas the rotating metric \textcolor{red}{(\ref{21})} covers all values of the spin parameter. The Fig.~\ref{horizon} shows the effect of the dark matter halo on the event horizon radius for $a=0.5M_{\rm BH}$ and $a=0.95M_{\rm BH}$ with different values of compactness parameter ${\cal C} = M/a_{0}$. As is clear, for a fixed value of spin parameter $a$, decreasing ${\cal C}$ causes the event horizon radius decreases so that for Kerr BH with ${\cal C } = 0$ it is smaller than that of rotating Cardoso BH.

\subsection{Electromagnetic properties of thin accretion disk}
\label{3-3-disk}
\begin{figure}[H]
\centering
\includegraphics[width=3.0in]{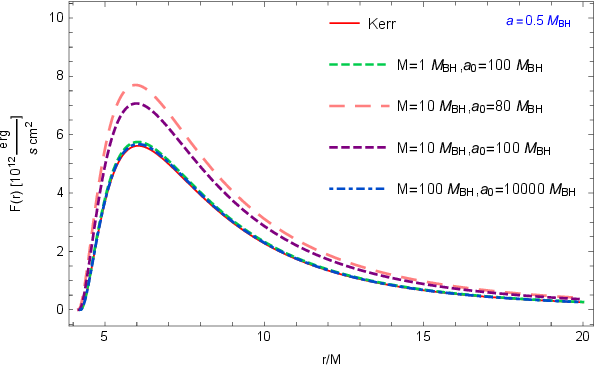}
\includegraphics[width=3.0in]{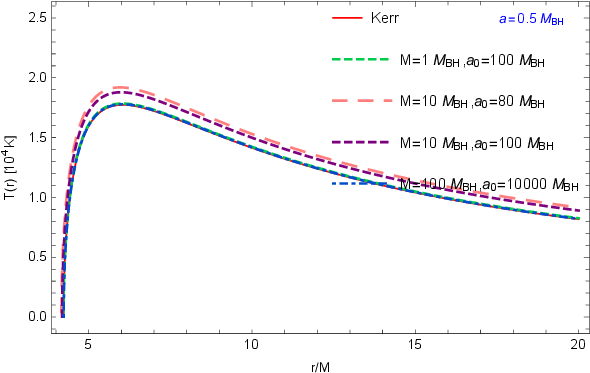}
\caption{\footnotesize The energy flux $F(r)$ from a disk (left panel) and the disk temperature $T(r)$ (right panel) around a rotating Cardoso BH with the mass accretion rate $\dot{M}=2\times10^{-6}M_{\odot}\rm yr^{-1}$, for different values of $M$ and $a_0$ with $a=0.5M_{\rm BH}$. In each panel the solid curve corresponds to the Kerr BH without dark matter.}
\label{flux}
\end{figure}

\begin{figure}[H]
\centering
\includegraphics[width=3.0in]{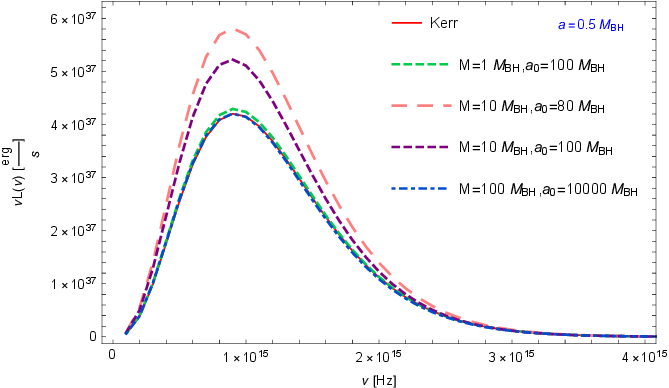}
\caption{\footnotesize The emission spectrum $\nu L(\nu)$ of the accretion disk around a rotating Cardoso BH with mass accretion rate $\dot{M}\sim10^{18}\rm g$ $ \rm s^{-1}$ and inclination $\gamma=0^{\circ}$, for different values of $M$ and $a_0$ with $a=0.5M_{\rm BH}$. The solid curve represents the disk spectrum for the Kerr BH without dark matter.}
\label{luminosity}
\end{figure}

Now, we are going to study the electromagnetic properties of accretion disk around the rotating Cardoso BHs, equation \textcolor{red}{(\ref{21})}, in the framework of the steady-state Novikov-Thorne model. First, we numerically calculate the specific energy ${\tilde{E}}$, the specific angular momentum ${\tilde{L}}$, and the angular velocity $\Omega$ of particles moving around the rotating Cardoso BHs. Then, substituting these results into equation \textcolor{red}{(\ref{11})} and equations \textcolor{red}{(\ref{12})}, \textcolor{red}{(\ref{13})}, \textcolor{red}{(\ref{14})}, we numerically obtain the ISCO radius, the radiant energy flux $F(r)$, the disk temperature $T(r)$, and the emission spectra $L(\nu)$ for the accretion disk around the central galactic rotating BHs immersed in Hernquist-type dark halo and plot the results in Fig.~\ref{flux} and Fig.~\ref{luminosity} for different values of compactness parameter. It also shows a comparison with the corresponding results for a Kerr
BH in the absence of dark matter. We see that the disks around rotating BHs at the center of Hernquist-type density distribution are much hotter and luminous than the Kerr BH and with increasing compactness parameter, the deviation from Kerr also increases.

In Tab.~\ref{t1}, we present the numerical values of $r_{\rm isco}$, the maximum values of energy flux, the disk's temperature, emission spectra and the radiative
efficiency of rotating Cardoso BHs for different values of compactness parameters, ${\cal C} = M/a_{0}$ for two spin parameters $a = 0.5M_{\rm BH}$ and $a = 0.9M_{\rm BH}$. Our analysis shows that for a given value of compactness parameter ${\cal C}$, the ISCO radius decrease with increasing the spin parameter $a$, and similarly, at a fixed value of spin parameter, the ISCO radius decrease with increasing the compactness parameter. Since the inner edge of disk in the Novikov-Thorne model lies at the ISCO radius, the decrease of ISCO radius leads to an increase of the disk surface, as well as the radiant energy flux, its temperature and luminosity in accordance with Fig.~\ref{flux} and Fig.~\ref{luminosity}.

\begin{figure}[H]
\centering
\includegraphics[width=3.0in]{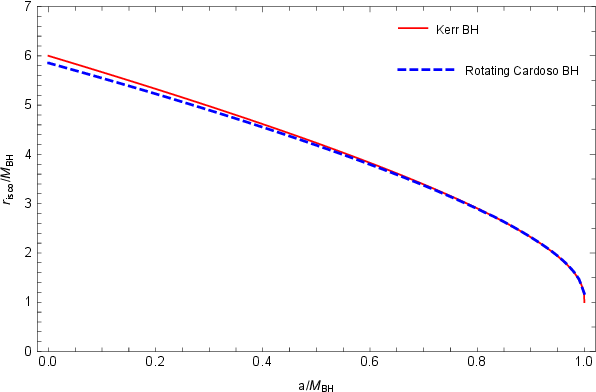}
\caption{\footnotesize The behavior of the ISCO radius as a function of the spin parameter in rotating Cardoso BH with $M=10 M_{BH}$ and $a_0=100 M_{BH}$ and in the Kerr BH for co-rotating orbits.}
\label{isco}
\end{figure}

In Tab.~\ref{t2}, we compare rotating Cardoso BH with the Kerr BH and static Cardoso BH. The comparison between rotating Cardoso BH and static Cardoso BH, as well as between rotating Cardoso BH and the Kerr BH, shows that the effect of the rotation parameter, $a$, in decreasing the ISCO radius is more prominent in comparison to the dark matter halo parameter. In Fig.~\ref{isco}, we also plot the behavior of the ISCO radius as a function of the spin parameter for rotating Cardoso BH and the Kerr BH. As is clear from the figure, dark matter does not play a significant role in decreasing the ISCO radius. In the other word, the rotating Cardoso BHs with a large spin are indistinguishable from the Kerr BH. Since the astrophysical BHs are expected to be highly spinning, it seems that the optical properties of accretion disks cannot be considered as a promising tool to distinguish a Kerr BH from a rotating BH immersed in Hernquist-type dark matter halo or even various types of dark matter halos.

\begin{table}[H]
\centering
\caption{ \footnotesize The numerical results for $r_{\rm isco}$, the maximum values of the radiant energy flux $F(r)$, temperature distribution $T(r)$, the emission spectra $\nu L(\nu)$, and the disk radiative efficiency around a rotating Cardoso BH for different values of compactness parameter. The first row corresponds to the Kerr BH without dark matter.}
\begin{tabular}{l l l l l l l l l l}
\hline\hline
$a/M_{\rm BH}$ & $M/M_{\rm BH}$ & $a_{0}/M_{\rm BH}$ & ${\cal C}$ & $r_{\rm isco}/M_{\rm BH}$ & $F_{\rm max}$$\times10^{12}$ &   $T_{\rm max}$$\times10^{4}$     & $\nu L(\nu)_{\rm max}$ $\times10^{37}$     & $\epsilon$
\\ [0.5ex]
\hline
\\
{0.5}
&-- & -- &-- &4.2330 & 5.6156 & 1.7740&4.2046&0.0821\\

&100 &  10000 &  0.01& 4.2329 & 5.6728 & 1.7785 & 4.2062&0.0906\\

&1 &100&0.01&4.2280& 5.7484 &1.7844&4.2957&0.0912\\

&10&100& 0.1&4.1857&7.0678&1.8790&5.2243&0.1665\\

&10&80&0.125&4.1632&7.7004&1.9197&5.6983&0.1856\\
\hline
\\
{0.9}
&-- & -- &-- &2.3209&7.8449 & 3.5012&1.4643$\times10$& 0.1557 \\

&100 &  10000 &  0.01& 2.3209& 7.8449& 3.5100 & 1.4643$\times10$&  0.1638\\

&1 &100&0.01&2.3208& 7.9415& 3.5119&1.4824$\times10$& 0.1642 \\

&10&100& 0.1&2.3200& 8.8907&3.6119&1.6596$\times10$& 0.2352\\

&10&80&0.125&2.3197&9.2346&3.6460&1.7237$\times10$& 0.2536\\
\hline\hline
\end{tabular}
\label{t1}
\end{table}

\begin{table}[H]
\centering
\caption{ \footnotesize The numerical results for $r_{\rm isco}$, the maximum values of the radiant energy flux $F(r)$, temperature distribution $T(r)$ and the emission spectra $\nu L(\nu)$ for a Kerr BH, rotating Cardoso BH and static Cardoso BH with ${\cal C}=0.1$.}
\begin{tabular}{l l l l l l l l }
\hline\hline
$type$&$a/M_{\rm BH}$ & $M/M_{\rm BH}$ & $a_{0}/M_{\rm BH}$ & $r_{\rm isco}/M_{\rm BH}$ & $F_{\rm max}$$\times10^{12}$ &   $T_{\rm max}$$\times10^{4}$     & $\nu L(\nu)_{\rm max}$$\times10^{37}$
\\ [0.5ex]
\hline
Rotating Cardoso BH&0.5&  10 &  100& 4.1857 & 7.0678&1.8790&5.2243 \\

Static Cardoso BH \redcite{1}&0&10&100&5.8552& 1.7867& 1.3323& 2.5348 \\

Kerr BH&0.5 & -- &-- &4.2330 & 5.6156 & 1.7740&4.2046  \\
\hline\hline
\end{tabular}
\label{t2}
\end{table}

\section{Conclusions}
The static, spherically symmetric BHs surrounded by a dark matter halo with a Hernquist-type density profile have been obtained by Cardoso et. al. in Ref.\redcite{Cardoso:2021wlq}. By studying the circular motion of test particles in accretion disks around static Cardoso BHs, we have numerically calculated the radius of the innermost stable circular orbits (ISCO) and the electromagnetic properties of such accretion disks in the framework of the Novikov-Thorne model in Ref. \redcite{Heydari-Fard:2024wgu}. Since the realistic BHs are expected to be rapidly rotating, in the present work by using the Newman-Janis algorithm, we first generalized the static BHs in Ref.\redcite{Cardoso:2021wlq} to the case of rotating BHs immersed in a Hernquist dark halo with zero radial pressure and non-zero tangential pressure. Then, by studying the circular motion of test particles in thin accretion disks, we have calculated the ISCO radius and the luminosity of BH's accretion disk. We also investigated the effects of spin parameter and halo compactness parameter on the ISCO radius and the disk properties and showed that for a given value of halo compactness parameter ${\cal C} = M/a_0$, the ISCO radius decreases with increasing the spin parameter, $a$, and similarly at a fixed value of $a$, the ISCO radius decreases with increasing ${\cal C} $. Comparing the results to that of Kerr BH in the absence of dark matter points out that the presence of dark matter does not play a significant role in decreasing the ISCO radius. Indeed, our analysis shows that the rotating Cardoso BHs with a large spin are indistinguishable from the Kerr BH. So, for astrophysical BHs with large spin parameter, it seems that the optical properties of accretion disks cannot be considered as a promising tool to distinguish a Kerr BH from a rotating BH immersed in Hernquist-type dark matter halo or even various types of dark matter halos.

\section*{Acknowledgements}
Malihe Heydari-Fard would like to thank the Iran National Science Foundation (INSF) and the Research Council of The University of Qom for financial support of project.

\end{document}